\documentstyle[a4,12pt,epsf]{article}
\epsfverbosetrue
\newcommand{\figurebox}[2]{\fbox{\vbox to#2in{\hbox to #1in{\hfil} \vfil}}}
\newcommand{\beq}{\begin{equation}}
\newcommand{\eeq}{\end{equation}}

\begin{document}

\begin{titlepage}

\begin{flushright}
Liverpool preprint: LTH 283\\
\today
\end{flushright}
\vspace{5mm}
\begin{center}
{\LARGE\bf Finite Size Effects and Scaling in Lattice $CP^{N-1}$}\\[1cm]

{\bf A.C.~Irving} and {\bf C.~Michael}\\[1cm]
DAMTP, University of Liverpool, Liverpool L69~3BX, UK

\end{center}
\vskip 2.0 true cm
\begin{abstract}

We present model predictions for the spectrum of $CP^{N-1}$ in a
periodic box and use them to interpret the strong finite size effects observed
in lattice simulations at medium values of $N$. The asymptotic scaling
behaviour of
alternative lattice actions is discussed along with some aspects of
multigrid algorithm efficiency.

\end{abstract}

\end{titlepage}

\paragraph{Introduction}\label{introduction}
There has been considerable recent interest in simulations of the
lattice $CP^{N-1}$
model \cite{JansenWiese,IM,HasenMeyer,Wolff,CampRossVic}.
Advances in the development of non-local Monte Carlo algorithms have
given added impetus to studies of the model at large $N$ where the
spectrum remains unconfirmed by analytic or numerical analysis. Conflicting
evidence has been presented for perturbative (two-loop) scaling at
currently accessible values of the bare coupling.

In this letter we analyse some key factors involved in such studies:
the dependence on lattice action, the efficiency of the algorithm and,
most crucially, contamination of finite
size effects which are particularly significant in the present model at
medium and large $N$. We demonstrate the latter effect by means of a
simplified analytic model and present some new data.

The simplest lattice action for the model containing an explicit
gauge field is
\begin{equation}
S=\beta/2\sum_{<xy>}[z_i^*(x)U(x,y)z_i(y) + h.c.]
\label{eq:SG}
\end{equation}
where $\beta = 2N/g^2$ and
${\bf z}(x)$ is a complex $N$-vector with unit norm.
$U(x,y)$ is a gauge connection between nearest neighbour sites
${x,y}$ and is an element of $U(1)$. This action, which was used in the
studies \cite{IM,CampRossVic}, corresponds in the classical continuum
limit to the Lagrangian
\begin{equation}
{\cal L} = {1\over{g^2}}(D_\mu{\bf z})^\dagger\cdot (D_\mu{\bf z})
\label{eq:LC}
\end{equation}
where the gauge field appearing in the covariant derivative
may be
related to the \lq matter\rq{} field ${\bf z}(x)$ via its equation of motion.
For computational convenience,
the lattice gauge variable can be integrated out to obtain an
effective $2D$ spin model
\begin{equation}
 Z = \int [d{\bf z}][d{\bf z}^*] \exp \{-\beta\tilde S({\bf z})\}
\label{eq:ZS}
\end{equation}
whose action is \cite{IM}
\begin{equation}
 \beta\tilde S= -\sum_{<xy>} \ln I_0(\beta |{\bf z}^*(x)\cdot{\bf z}(y)|)
\label{eq:SS}
\end{equation}

The alternative action
\begin{equation}
  S_Q= -2\kappa\sum_{<xy>} |{\bf z}^*(x)\cdot{\bf z}(y)|^2
\label{eq:SQ}
\end{equation}
can be shown to have the same classical continuum limit as (\ref{eq:SG})
with the identification of bare couplings $\kappa = N/2g^2$.
This latter action was studied
by \cite{JansenWiese,HasenMeyer,Wolff}.

At large $N$, an effective Lagrangian may be derived \cite{dAdda,Witten}
which corresponds to $N$ identical charged scalar fields subject to a
(confining) $U(1)$ gauge potential
\begin{equation}
 {\cal V}({\bf r}) = g_f^2|{\bf r}| \quad
\label{eq:V}
\end{equation}
where the effective coupling is related to the mass of the charged
scalar $m_f$
\begin{equation}
 g_f = m_f\sqrt{{12\pi}\over{N} \quad.}
\label{eq:gf}
\end{equation}
In the large $N$ limit, the low-lying spectrum can be estimated in a
non-relativistic approximation via a Schr\"odinger equation for the
two-body relative motion in the potential ${\cal V}({\bf r})$.
\cite{Haber}.
For medium $N$ ($4\dots 8$), a previous lattice analysis showed that the
predictions of this simple model are at best unreliable \cite{IM}. The
lowest state observed was indeed the adjoint, but there was little sign of
the next predicted state, an $SU(N)$ singlet. It was concluded that
either the lattice singlet operators used were inefficient or that the
breakdown of the non-relativistic model approximation at these $N$
values was to blame.

\paragraph{Model for finite size effects}
All analyses of $CP^{N-1}$ have noted strong finite-size effects.
Detailed knowledge of the $N$-dependent potential allows one to estimate
the size of lattice required to be free of such effects \cite{IM}. Most
analyses have required $L/\xi\geq 6$ for this reason. In fact, one can
detect the effects of working in a periodic box even above this limit
especially at large values of $N$ where the flat potential leads to a
weakly confined sytem and hence broad wave functions \cite{IM,CampRossVic}.

It is interesting to use the (albeit imperfect) non-relativistic model
described in the first section to investigate these effects more
quantitatively. The potential implied by eqns (\ref{eq:V}, \ref{eq:gf}) when
made periodic with period $L$ gives rise to a band rather than discrete
spectrum (see any Quantum Mechanics text for the treatment of the
Kronig-Penny potential \cite{Merz}). The discrete eigenvalues determined
by the zeroes of Airy functions \cite{Haber} give way to allowed bands
whose limits are determined by combinations of Airy functions
\begin{equation}
\Delta\bar\Delta\cos(\kappa L) = (B^\prime A + A^\prime B)
(\bar B^\prime\bar A + \bar A^\prime\bar B)
-2AA^\prime\bar B\bar B^\prime
-2BB^\prime\bar A\bar A^\prime \quad .
\label{eq:cos}
\end{equation}
Here, $\kappa$ is some real parameter and $L$ is the period of the potential:
the spatial extent of the lattice in our application. Also
\begin{equation}
\Delta=B^\prime A - A^\prime B,\qquad\bar\Delta=\bar B^\prime\bar A -
\bar A^\prime\bar B
\label{eq:delt}
\end{equation}
where $A\equiv Ai(-b)$ and $B\equiv Bi(-b)$ are the standard Airy functions,
with $A^\prime$, $B^\prime$ the corresponding derivatives.
Similarly $\bar A\equiv Ai(aL/2-b)$, and so on,
where the parameters $a$, $b$
are related to the potential slope $2V_0/L$, the required band energy $E$
and the reduced
mass of the two-particle system $\mu=m_f/2$:
\begin{equation}
a=\bigl({{4\mu V_0}\over L})^{1/3}, \qquad b/a={{EL}\over{2V_0}}\quad .
\label{eq:ab}
\end{equation}
In the present application, $V_0=g_f^2L/2$. All scales in the
model are set by the confined, and so unobservable, fundamental mass
$m_f$. To compare predictions with data (see below) the relationship to
the physical bound state mass must be used.

In Fig 1, is shown the band spectra for  $N=4$ and $8$ in units of $m_f$
as a function
of $1/L$. These have been checked against the results of numerical
solutions of the corresponding Sturm-Liouville problem. Using the
numerical approach we have also been able to replace the model potential
(\ref{eq:V}) by that measured in lattice simulations ~\cite{IM}. The
qualitative results remained similar to those seen in Fig 1. The
spectrum at $1/L=0$ agrees with that of Haber et al. ~\cite{Haber}. i.e.
a low-lying positive parity adjoint state followed by well separated
negative parity and further excited states. As the size of the periodic
box decreases in physical units, a band structure develops implying that
zero momentum correlators in lattice simulations will no longer
be dominated by a single isolated state.
At increasing Euclidean time separation, the contribution from the lower
levels within the lowest band will dominate so that the effective mass
$m_{eff}(t)$ defined by
\begin{equation}
C(t)=\hbox{const.} \times [
\exp(-m_{eff}t)+\exp(-m_{eff}(L-t)) ]
\label{eq:meff}
\end{equation}
shows a drop-off in $t$ rather than the plateau characteristic of an
isolated state. One may construct simple models for the overlap of the band
states in Fig 1 with the
measurement operator employed in $C(t)$
which demonstrate the transition from $m_{eff}(t)\approx m_0$ at moderate
Euclidean time separations to $m_{eff}(t)\approx m_0-\delta/2$ at large
separations. Here, $m_0$ is the centre of the energy band for some
lattice size and $\delta$ its width and all units set by $m_f=1$.

Since the non-relativistic model is known to be quantitativley unreliable
{}~\cite{IM} it is difficult to make other than qualitative statements on
the implications of this effect. As an example, for $N=4$ and $\kappa=2.9$
(action (\ref{eq:SQ})), $\xi/a=18.5$. According to the non-relativistic
treatment ~\cite{Haber} for $N=4$
\begin{equation}
m=1/\xi=(2+4.54)m_f
\label{eq:mNR}
\end{equation}
Thus, a lattice of
size $L/a=128$, say, corresponds to $L=1.06m_f^{-1}$. According to Fig 1,
this implies a very large broadening effect indeed. Of course, we know
that the system is actually relativistic since eq.(\ref{eq:mNR}) implies
$v/c\approx 0.95$ so model predictions are unreliable. On the other
hand, the string tension $\sigma$ implied by the large $N$ model ($\sigma
/m^2=3\pi/6.54^2=0.22$) is
comparable with that measured ($0.34(2)$) ~\cite{IM} when expressed as
a dimensionless ratio.

Hasenbusch and Meyer ~\cite{HasenMeyer} have presented  correlator data
at very large time separation, which the authors claim is essential to
obtain the true mass-gap in the presence of other states. While this is
in general true, the above model demonstrates that for
systems bound loosely by a shallow potential such as large $N$ $CP^{N-1}$
the effect of periodic boundaries could be such that large time
separations isolate the wrong state i.e. not that which is physically
relevant in the bulk, continuum limit. In the absence of detailed
published data for $m_{eff}(t)$ we have investigated this point by studying
$CP^3$ and $CP^7$ at weak coupling on small ($64^2$) and large
($128^2$) lattices. Fig 2
shows typical results. There is some small but significant effect
apparent, particularly for $N=8$ where the potential is flatter. Note
that the errors shown are statistically correct at each time separation
but are correlated in Euclidean time. Clearly the effect is considerably
smaller than the naive non-relativistic model (Fig 1) predicts.
Nonetheless, we would suggest that there is a significant source of systematic
error which must be allowed for in future high statistics analyses.
To be free of finite size effects, it would seem prudent to study
systems for which ~\cite{IM}
\begin{equation}
L/\xi>cN
\label{eq:Lxi}
\end{equation}
where $c$ is of the order of $2$.

For much smaller lattices, one sees a reversal to the more usual
finite size effect of {\it raised} effective masses (reduced correlation
lengths) ~\cite{CampRossVic}.

\paragraph{Scaling properties}
Analyses using the quadratic action (\ref{eq:SQ}) have shown conflicting
evidence for scaling of the mass gap (adjoint state)
\cite{JansenWiese,HasenMeyer}. The first study with relatively low
statistics and covering the lattice correlation range $\xi/a<30$ showed
some evidence for scaling (constant ratio of dimensionful quantities)
and for asymptotic scaling of the mass gap itself \cite{Hikami,Brezin}
\begin{equation}
  ma=(m/\Lambda_N)({1\over{g^2}})^{2\over N}
\exp\{-{{2\pi}\over{g^2}}\}
\label{eq:ma}
\end{equation}
This analysis \cite{JansenWiese} relied largely on a local
Metropolis algorithm. However, a
multigrid approach enabled Hasenbusch and Meyer \cite{HasenMeyer} to
achieve larger statistics and to probe larger correlation lengths  ($\leq
77$). Clear evidence of non-asymptotic scaling was presented.

Irving and Michael used the gauge-explicit action (\ref{eq:SG},\ref{eq:SS})
for the models with $4\leq N\leq 8$
and, using a hybrid local/cluster algorithm, presented data showing evidence
of scaling (ratio of topological charge and mass-gap) and for asymptotic
scaling of the mass-gap for $g^2 \leq 1.1$. A subsequent
analysis ~\cite{CampRossVic}
using a hybrid over-relaxation/heat bath algorithm on the same action
with explicit gauge fields has provided further convincing evidence that both
scaling and asymptotic scaling properties are superior for this action
when compared to the quadratic action (\ref{eq:SG}).

We have used a variant of the multigrid algorithm of Hasenbusch and
Meyer \cite{HasenMeyer} to compare the asymptotic scaling in the $N=4$ model
(\ref{eq:ma}) for both actions using the {\it same} analysis techniques. The
$t$-dependent effective mass-gap was extracted from zero-momentum correlators
using eqn. \ref{eq:meff}. The errors on the
$t$-independent values (see discussion above) were extracted using
bootstrap techniques. Typically 20000 multigrid V-cycles with
measurements every second cycle were made at each coupling.
As can be seen from
table~\ref{tab:scvl}, the corresponding
values of $m/\Lambda$ are not constant in $1/g^2$ but are significantly more
slowly varying for the explicit gauge action $S$.
If one parameterises
the the scaling violation by
\begin{equation}
  m/\Lambda \sim (\xi/a)^{-z_s}
\label{eq:ScViol}
\end{equation}
then the data presented in table ~\ref{tab:scvl} yield
$z_s=.18\pm.02$
($.07\pm .02$) for the quadratic (explicit gauge) action.
\begin{table}
\begin{center}
\begin{tabular}{|l|l|l|l|r|r|}
\hline
\multicolumn{3}{|c|}{Quadratic action} &
\multicolumn{3}{|c|}{Explicit gauge action}\\\hline
\multicolumn{1}{|c}{$1/g^2$}  &
\multicolumn{1}{|c}{$\xi/a$} &
\multicolumn{1}{|c}{$m/\Lambda$} &
\multicolumn{1}{|c}{$1/g^2$} &
\multicolumn{1}{|c}{$\xi/a$} &
\multicolumn{1}{|c|}{$m/\Lambda$} \\\hline
1.25  & 4.48(8) & 514(9) & 0.909  & 3.1(1) & 102(3) \\
1.35  & 8.7(3)  & 478(12) & 0.977 & 4.9(1) & 97(2) \\
1.4   & 12.8(3) & 436(11) & 1.000 & 5.3(1) & 98(3) \\
1.45  & 18.5(3) & 406(8)  & 1.111 & 10.8(3) & 95(3) \\
1.5   & 30.0(2) & 337(15) & 1.190 & 18.5(7) & 88(3) \\\hline
\end{tabular}
\caption{{\it The scaled adjoint mass $m/\Lambda$ in lattice
$CP^3$ using the quadratic action and the explicit gauge
action (see text). Also shown are the corresponding bare couplings
and correlation lengths in lattice units ($\xi=1/m$).}
\label{tab:scvl}
}
\end{center}
\end{table}

\paragraph{Algorithms}
The data presented above was obtained using an adaptation of the
multigrid algorithm proposed by Hasenbusch and Meyer ~\cite{HasenMeyer}.
The particular improvement tested was to use random $SU(2)$ subgroups
rather than $U(1)$ and to employ a different subgroup and rotation at
each level of the V-cycle. Using this, we made careful measurements of
the decorrelation dynamical exponent $z$ using the spin susceptibility
as a good measure for the self-consistent autocorelation time. For
example, we found
values close to $1.0$ for the quadratic action (\ref{eq:SQ})
\begin{equation}
z_{(N=4)}=1.05\pm.08,\qquad
z_{(N=8)}=1.12\pm.10
\label{eq:z}
\end{equation}
i.e. considerably larger than those reported by ~\cite{HasenMeyer} but
comparable with those reported by Wolff ~\cite{Wolff} who used an
overrelaxation algorithm. When we used a fixed $U(1)$ rotation for each
block within the V-cycle, in the spirit of ~\cite{HasenMeyer} the
decorrelation performance of our algorithm was degraded.
The multigrid performance on the
action (\ref{eq:SS}) was similar: the dynamical exponent for $CP^3$ was
$1.00\pm .07$.
Of course, all these algorithms offer considerable improvement over the
local Metropolis, single cluster, and hybrid algorithms used in earlier
work ~\cite{JansenWiese,IM}.
\paragraph{Conclusions}
We have presented a model for non-trivial finite size effects in
$CP^{N-1}$ and new data relevant to this, to the question of precocious
scaling and to the efficiency of multigrid algorithms. Effects ascribable
to periodicity as predicted by the simple model have been demonstrated but
at a much lower level. We have confirmed earlier evidence
{}~\cite{IM,CampRossVic} that the explicit gauge action for lattice
$CP^{N-1}$ has superior scaling properties to the more commonly used
quadratic action. Finally, we reported measurements of the dynamical
exponent for a multigrid algorithm which represent a considerable
improvement on local algorithms but which fall short of other reported
data for similar algorithms ~\cite{HasenMeyer}.

\paragraph{Acknowledgements}

This research is supported by the UK Science and Engineering Research Council.

\newpage
\paragraph{Figure captions}\label{figs}
\begin{enumerate}
\item The low-lying part of the predicted energy spectrum due to a
periodic potential. The energy bands allowed by (\ref{eq:cos}) are
indicated by the shaded regions bounded by solid(dashed) lines for
$N=4(8)$. Units are set by $m_{eff}=1$.

\item The effective mass defined by eqn. \ref{eq:meff} for
(a) $N=4$ at $\kappa=2.9$
and (b) $N=8$ at $\kappa=5.5$ The crosses (circles) denote data on
$128^2(64^2)$ lattices. The solid line is the asymptotic mass gap used in
the scaling analyses.
\end{enumerate}
\end{document}